\documentclass[aps,prd,preprintnumbers,superscriptaddress,nofootinbib,showpacs]{revtex4}
\usepackage[dvips]{graphicx}
\usepackage{bm,latexsym,amsmath,amssymb,amsfonts,mathrsfs}
\usepackage{color}
\input{colordvi.tex}
\newcommand*{\D}{{\rm d}}
\newcommand*{\mpl}{\widetilde{M}_{\rm Pl}}
\begin{document}

\title{Breaking of Vainshtein screening in scalar-tensor theories beyond Horndeski}

\author{Tsutomu~Kobayashi}
\email[Email: ]{tsutomu"at"rikkyo.ac.jp}
\affiliation{Department of Physics, Rikkyo University, Toshima, Tokyo 175-8501, Japan
}

\author{Yuki~Watanabe}
\email[Email: ]{watanabe"at"resceu.s.u-tokyo.ac.jp}
\affiliation{Research Center for the Early Universe (RESCEU), Graduate School of Science,
The University of Tokyo, Tokyo 113-0033, Japan
}

\author{Daisuke~Yamauchi}
\email[Email: ]{yamauchi"at"resceu.s.u-tokyo.ac.jp}
\affiliation{Research Center for the Early Universe (RESCEU), Graduate School of Science,
The University of Tokyo, Tokyo 113-0033, Japan
}

\begin{abstract}
The Horndeski theory of gravity is known as the most general scalar-tensor theory
with second-order field equations. Recently, it was demonstrated
by Gleyzes {\em et al.} that
the Horndeski theory can further be generalized in such a way that although
field equations are of third order, the number of propagating degrees of
freedom remains the same. 
We study small-scale gravity
in the generalized Horndeski theory, focusing in particular on an impact of
the new derivative interaction beyond Horndeski on the Vainshtein screening mechanism.
In the absence of the quintic Galileon term and its generalization,
we show that
the new interaction does not change the qualitative behavior of
gravity outside and near the source:
the two metric potentials coincide, $\Phi = \Psi \;(\sim r^{-1})$,
while the gravitational coupling is given by the cosmological one
and hence is time-dependent in general.
We find, however, that the gravitational field inside the source
shows a novel behavior due to the interaction beyond Horndeski:
the gravitational attraction is not determined solely from the enclosed mass
and two potentials do not coincide, indicating breaking of the screening mechanism.
\end{abstract}

\preprint{RUP-14-15, RESCEU-47/14}
\maketitle

\section{Introduction}

A number of modified gravity theories have been proposed so far
as alternatives to dark energy~\cite{Clifton:2011jh}.
Such theories are typically described by scalar-tensor gravity,
in which a scalar propagating degree of freedom is introduced
in addition to the gravitational wave ones.
A single scalar and two tensor modes imply that
equations of motion for the scalar
field 
and the metric are of second order.
The Horndeski theory of gravity~\cite{Horndeski:1974wa}, the most general scalar-tensor theory
with second-order field equations, is therefore quite useful
for a comprehensive study of modified gravity.

Recently, it has been noticed that
the Horndeski theory can be generalized without introducing
any degrees of freedom other than the single scalar and two tensor modes~\cite{GLPV,Lin:2014jga,GLPV2}.
Going beyond the Horndeski class, one inevitably obtains higher order field equations
since the Horndeski Lagrangian has been proven to be the most general one
with second-order field equations.
The trick is 
to introduce a preferred time slicing on which
the necessary number of initial conditions remains the same as that of the Horndeski theory.
The preferred slicing is given by the unitary gauge
in which the scalar field $\phi$ is spatially homogeneous,
and away from the unitary gauge the field equations
are of higher order in general.
One can thus obtain a more general framework to study modified gravity
based on scalar-tensor theories beyond Horndeski.
Aspects of scalar-tensor theories beyond Horndeski have been investigated
in Refs.~\cite{Kase:2014baa,Kase:2014yya,Fasiello:2014aqa}.
For a further generalization of the work of~\cite{GLPV},
see Refs.~\cite{Gao, Gao2}.

The scalar degree of freedom in modified gravity is supposed to accelerate the
current cosmic expansion, while care must be taken
because the scalar-mediated force would persist down to small scales
where a deviation from general relativity (GR) is strongly
constrained, {\em e.g.}, by the solar-system tests.
A screening mechanism for the scalar force on small scales is therefore required to be
incorporated in modified gravity.
The Vainshtein mechanism~\cite{Vainshtein:1972sx} is one such screening mechanism
that operates in theories possessing second derivatives of the scalar
in the Lagrangian. The Galileons~\cite{Nicolis:2008in},
DGP gravity~\cite{Dvali:2000hr}, and massive gravity~\cite{Fierz:1939ix,deRham:2010kj}
are known to exhibit Vainshtein screening in the vicinity of
a gravitational source~\cite{Babichev:2013usa}.
The original Horndeski Lagrangian can be written equivalently in
the form of the generalized Galileon~\cite{Deffayet:2011gz,GG},
and this Galileon-like structure allows us to study
the Vainshtein mechanism in a generic manner.
Several works have been carried out along this
direction: the generic conditions for the Vainshtein mechanism to work
were derived and some observational implications were suggested
in~\cite{Kimura:2011dc, Narikawa:2013pjr},
and the stability of the screened solutions was discussed in~\cite{KNT}.
See also Refs.~\cite{DeFelice:2011th,Kase:2013uja} for related discussions.
The purpose of this paper is to extend those previous works and
to study how the scalar-tensor Lagrangian beyond Horndeski
changes the behavior of small-scale gravitational fields.

This paper is organized as follows. In the next section
we review how the Horndeski theory can be generalized
while maintaining the number of propagating degrees of freedom~\cite{GLPV}.
In Sec.~III we derive the cosmological background equations
in the generalized Horndeski theory. In Sec.~IV,
the effective Lagrangian that governs small-scale gravity
on  the cosmological background
is obtained taking into account all relevant nonlinear terms.
We then discuss
the possible impact of the new terms arising from the Lagrangian beyond Horndeski:
first, we investigate linear density perturbations in Sec.~V,
and next we study the Vainshtein mechanism for spherical overdensities
in Sec.~VI.
We draw our conclusions in Sec.~VII.

\section{Scalar-tensor theories beyond Horndeski}

The Horndeski theory, {\em i.e.},
the most general scalar-tensor theory with {\em second-order} field equations,
is described by 
\begin{eqnarray}
S_{\rm H}=\int\D t\D^3\mathbf{x}\sqrt{-g}L_{\rm H}[g_{\mu\nu},\phi ].
\end{eqnarray}
The action controls the dynamics of gravity and the scalar field, $\phi$, and its explicit expression is given by
\begin{eqnarray}
\int\D t\D^3\mathbf{x}\sqrt{-g}L_{\rm H}=\int\D t\D^3\mathbf{x}\sqrt{-g}\sum_{a=2}^5L_a,
\end{eqnarray}
where
\begin{eqnarray}
L_2&:=&G_2(\phi, X),
\\
L_3&:=&-G_3(\phi, X)\Box\phi,
\\
L_4&:=&G_4(\phi,X)R+G_{4X}\left[(\Box\phi)^2
-(\nabla_\mu\nabla_\nu\phi)^2\right],
\\
L_5&:=&G_5(\phi,X)G_{\mu\nu}\nabla^\mu\nabla^\nu\phi
-\frac{1}{6}G_{5X}\left[
(\Box\phi)^3-3\Box\phi(\nabla_\mu\nabla_\nu\phi)^2+2(\nabla_\mu\nabla_\nu\phi)^3\right].
\end{eqnarray}
and each $G_a$ is an arbitrary function of $\phi$ and
$X:= -g^{\mu\nu}\partial_\mu\phi\partial_\nu\phi/2$.
Throughout the paper,
subscripts $X$ and $\phi$ are understood as differentiation,
{\em e.g.}, $G_{aX}:= \partial G_a/\partial X$.

In the unitary gauge, $\phi=\phi(t)$, we have $X=\dot\phi^2/(2N^2)$
with $N$ being the Lapse function,
and hence a function of $\phi$ and $X$ can be regarded as that of $t$ and $N$.
Here and hereafter a dot ( $\dot{}$ ) indicates a derivative with respect to $t$.
This leads to the following unitary gauge description of the Horndeski theory:
\begin{eqnarray}
L_{\rm H}&=&A_2(t, N)+A_3(t, N)K+
A_4(t, N)\left(K^2-K_{ij}^2\right)+B_4(t, N)R^{(3)}
\nonumber\\&&
+A_5(t,N)\left(K^3-3KK_{ij}^2+2K_{ij}^3\right)
+B_5(t,N)K^{ij}\left(R_{ij}^{(3)}-\frac{1}{2}g_{ij}R^{(3)}\right)
\label{Uni-L}
\end{eqnarray}
with
\begin{eqnarray}
A_4=-B_4-N\frac{\partial B_4}{\partial N},
\quad
A_5=\frac{N}{6}\frac{\partial B_5}{\partial N},\label{H-relation}
\end{eqnarray}
where $K_{ij}$ and $R_{ij}^{(3)}$ are the extrinsic and intrinsic curvature tensors
of the spatial hypersurfaces, respectively.

The central idea of Ref.~\cite{GLPV} is that
one can maintain the number of the propagating degrees of freedom
even if $A_4$ and $A_5$ are taken to be arbitrary functions of $t$ and $N$
rather than the functions constrained as Eq.~(\ref{H-relation}).
This generalization does not change the essential structure
of the equation of motion for $N$: it does not give the
constraint among the dynamical variables, giving rise to
one scalar degree of freedom on top of the usual gravitational wave modes.
Ungauging the unitary gauge description~(\ref{Uni-L}) with arbitrary $A_4$ and $A_5$,
one obtains the generalized Horndeski theory
with the following new terms~\cite{GLPV}:
\begin{eqnarray}
\widetilde{L}_4&:=&
F_4(\phi, X)
\biggl\{
\nabla^\mu\phi\nabla^\nu\phi\nabla_\mu\nabla_\nu\phi\Box\phi
-\nabla^\mu\phi\nabla_{\mu}\nabla_{\lambda}\phi\nabla^\nu\phi\nabla_{\nu}\nabla^\lambda\phi
+X\left[(\Box\phi)^2
-(\nabla_\mu\nabla_\nu\phi)^2\right]
\biggr\}
\notag\\
&=&-\frac{1}{2}F_4(\phi ,X)
\epsilon^{\mu\nu\alpha\beta}
\epsilon_{\mu'\nu'\alpha'\beta}\nabla^{\mu'}\phi\nabla_\mu\phi
\nabla^{\nu'}\nabla_\nu\phi\nabla^{\alpha'}\nabla_\alpha\phi,\label{eq:tilde L4}
\\
\widetilde{L}_5&:=&
F_5(\phi,X)
\biggl\{
(\Box\phi)^2\nabla^\mu\phi\nabla^\nu\phi\nabla_{\mu}\nabla_\nu\phi
-2\Box\phi\nabla_\mu\phi\nabla^{\mu}\nabla^{\nu}
\phi\nabla_{\nu}\nabla_{\lambda}\phi\nabla^\lambda\phi
\nonumber\\&&
\quad\quad\quad
-(\nabla_\mu\nabla_\nu\phi)^2\nabla^\rho\phi\nabla^\lambda\phi\nabla_\rho\nabla_\lambda\phi
+2\nabla_\mu\phi\nabla^\mu\nabla^\nu\phi\nabla_{\nu}\nabla_\rho\phi
\nabla^\rho\nabla^\lambda\phi\nabla_\lambda\phi
\nonumber\\&&
\quad\quad\quad
+\frac{2}{3}X\left[
(\Box\phi)^3-3\Box\phi(\nabla_\mu\nabla_\nu\phi)^2+2(\nabla_\mu\nabla_\nu\phi)^3\right]
\biggr\}
\notag\\
&=&-\frac{1}{3}F_5(\phi ,X)\epsilon^{\mu\nu\alpha\beta}
\epsilon_{\mu'\nu'\alpha'\beta'}\nabla^{\mu'}\phi\nabla_\mu\phi
\nabla^{\nu'}\nabla_\nu\phi\nabla^{\alpha'}\nabla_\alpha\phi\nabla^{\beta'}\nabla_\beta\phi,\label{eq:tilde L5}
\end{eqnarray}
where $\epsilon^{\mu\nu\alpha\beta}$ is the totally antisymmetric Levi-Civita tensor density of weight $-1$.
Even though adding these terms to the Lagrangian results in
{\em third} derivatives in the field equations,
the number of propagating degrees of freedom remains unchanged.
This fact has recently been verified by the detailed Hamiltonian analysis~\cite{Lin:2014jga, GLPV2}.

In this paper, we work in the generalized Horndeski theory
with the matter sector minimally coupled to gravity, the action of which
is given by
\begin{eqnarray}
S=\int\D t\D^3\mathbf{x}\sqrt{-g}L_{\rm grav}[g_{\mu\nu},\phi ]+S_{\rm matt}[g_{\mu\nu},\psi,A_{\mu},\cdots ]
=\int\D t\D^3\mathbf{x}\sqrt{-g}\left( L_{\rm H}+\widetilde{L}_4+\widetilde{L}_5\right) +S_{\rm matt}.
\label{eq:action}
\end{eqnarray}
The matter sector is composed of the standard model particles and cold dark matter, which form structures such as galaxies, galaxy clusters, and halos.

Before closing this section, it is worth noting that
one can further generalize the above theory without
introducing any extra degrees of freedom.
Indeed, following the spirit of the effective field theory approach
it is possible to add various terms constructed from $K_{ij}$, $R_{ij}^{(3)}$, and
$a_i:=\partial_i\ln N$ having the coefficients
that are functions of $t$ and $N$~\cite{Gao}.
This construction
yields derivatives higher than three in the field equations in general,
while not increasing the number of propagating degrees of freedom.
The resultant theory includes, for example, versions of Ho\v{r}ava-Lifshitz
gravity~\cite{Horava1, Horava2} as specific cases
because now the combination of the form $K_{ij}^2-\lambda K^2$
is allowed.
Nevertheless, in this paper
we focus on the theory introduced by Gleyzes {\em et al.}~\cite{GLPV}
as it can be regarded as the simplest extension of the Horndeski theory.

\section{Background evolution}

Let us begin with investigating the
cosmological background equations.
The metric and the scalar field are
spatially homogeneous and isotropic,
$g_{\mu\nu}\D x^\mu \D x^\nu =-N^2(t)\D t^2+a^2(t)\delta_{ij}\D x^i\D x^j$ and $\phi=\phi(t)$.
One can derive the field equations by varying the action
with respect to $N$, $a$ and $\phi$, and then by setting $N=1$.
The resultant equations are
\begin{eqnarray}
&&{\cal E}_{\rm H}+12H^2X^2\left(5F_4+2XF_{4X}\right)
-8H^3X^2\dot\phi\left(7F_5+2XF_{5X}\right)=-\rho,\label{00b}
\\
&&{\cal P}_{\rm H}-4X\left[\left(3H^2+2\dot H\right)XF_4+4H\dot X F_4+2HX\dot XF_{4X}
+2HX\dot\phi F_{4\phi}\right]
\nonumber\\&&
\quad
+8HX^2\left[
2\left(H^2+\dot H\right)\dot\phi F_5+5H\ddot\phi F_5+2HX\ddot\phi F_{5X}+2HXF_{5\phi}
\right]=-p,\label{ijb}
\end{eqnarray}
and $\dot J + 3HJ = P_\phi$ with
\begin{eqnarray}
J=J_{\rm H}+24H^2 X\dot\phi F_4+12H^2X^2\dot\phi F_{4X}
-40H^3X^2F_5-16H^3X^3F_{5X}
\end{eqnarray}
and
\begin{eqnarray}
P_\phi= P_{{\rm H}\phi}+12H^2X^2F_{4\phi}-8H^3X^2\dot\phi F_{5\phi}.
\end{eqnarray}
We have included the energy density and pressure of usual matter
in the right-hand sides of Eqs.~(\ref{00b}) and~(\ref{ijb}).
Here, ${\cal E}_{\rm H}$, ${\cal P}_{\rm H}$, $J_{\rm H}$, and
$P_{{\rm H}\phi}$
depend only on the Horndeski functions $G_a$, and their explicit forms
can be found in the literature~\cite{GG}.
(They are replicated in the appendix for the reader's convenience.)
One notices that ${\cal E}_{\rm H}$ is at most cubic in $H$
and no higher order terms are generated from $\widetilde{L}_4$
and $\widetilde{L}_5$ since the Lagrangian beyond Horndeski
is at most cubic in $K_{ij}$; the Friedmann equation is still
at most cubic in $H$ in the generalized Horndeski theory.

At sufficiently early times, we assume that the highest-order terms
in $H$ are dominant in the Friedmann equation and determine the expansion law of the Universe.
Note that the coefficient of the cubic term is composed only of $G_5$ and $F_5$.
Therefore, in theories with $G_5=0$ and $F_5=0$ we have the
early-time behavior as\footnote{To be precise, such a behavior
is true in models with $G_{5X}=F_{5X}=0$ since terms
with $G_5=G_5(\phi)$ and $F_5=F_5(\phi)$ can be integrated by parts and absorbed
into those with $G_4(\phi,X)$ and $F_4(\phi,X)$.}
\begin{eqnarray}
3H^2\approx 8\pi G_{\rm cos}\rho,
\end{eqnarray}
where
\begin{eqnarray}
8\pi G_{\rm cos}=\left[2G_4-8X\left(G_{4X}+XG_{4XX}\right)
-4X^2\left(5F_4+2XF_{4X}\right)\right]^{-1},
\end{eqnarray}
unless a tracker scaling solution is realized.
We thus see that in this case
the scalar-field contribution to the cosmic expansion
is screened at early times
except for the time-dependence of $G_{\rm cos}$.
At late times the other terms in the Friedmann equation
are assumed to become larger than the matter energy density to be
responsible for the accelerated expansion.

\section{Effective theory on small scales}

For the purpose of investigating the Vainshtein screening mechanism on small scales,
we construct an effective theory for small fluctuations
sourced by a matter overdensity $\delta$,
within which galaxies and associated objects are essentially frozen in their present-day configurations.
To do so, let us consider quasi-static perturbations on a cosmological background.
Working in the Newtonian gauge, we write the (dimensionful) metric perturbations,
the perturbed scalar field, and the nonrelativistic matter energy density as
\begin{eqnarray}\label{eq:metric_perturb}
&&
g_{\mu\nu}\rightarrow g_{\mu\nu}(t)+\frac{h_{\mu\nu}(t,\mathbf{x})}{\mpl},
\quad
h_{00}=-2\Phi(t,\mathbf{x}),
\quad
h_{ij}=-2a^2(t)\Psi(t,\mathbf{x}) \delta_{ij},
\\
&&
\phi\rightarrow \phi (t)+\pi(t, \mathbf{x}),
\end{eqnarray}
and
\begin{eqnarray}
\rho\rightarrow \rho (t)\left[ 1+\delta (t,\mathbf{x})\right],
\end{eqnarray}
where $\mpl$ is a mass scale which
is assumed to be of order of the Planck mass.

To study the quasi-static behavior of
those perturbations on small scales,
we expand the action~(\ref{eq:action}) in terms of
$h_{\mu\nu}$ and $\pi$ under the following rules.
We employ the quasi-static approximation, so that
$(\nabla\varepsilon)^2 \gg (\partial_t \varepsilon)^2$ in the action,
where $\varepsilon$ stands for any of $\Phi$, $\Psi$, and $\pi$,
and $\nabla$ denotes a spatial derivative.\footnote{Note that since we assume $(\partial_t\varepsilon)^2 \sim \max[H^2\varepsilon^2, M^2\varepsilon^2]$, where $M^{-1}$ is the Compton wavelength of the perturbations, the mass term is subleading and in particular screening due to $\pi$'s mass, {\it i.e.}, the chameleon mechanism~\cite{Khoury:2003aq} cannot take place. The size of the overdensity region one considers must then be much smaller
than the Compton wavelength $M^{-1}$.}
The matter overdensity $\delta$ is assumed to be of ${\cal O}(\nabla^{2}\Phi)$.
Expanding the action gives rise to terms with second derivatives acting on $\varepsilon$.
In particular, we will have many nonlinear terms of the form
$(\nabla\varepsilon)^2(\nabla^{2}\varepsilon)^n$
or equivalently $\varepsilon(\nabla^{2}\varepsilon)^{n+1}$, and also $\dot\varepsilon(\nabla^{2}\varepsilon)^{n}$
($n=1,2,\cdots$).
We keep all such terms because they can be larger than the quadratic terms below a certain scale.
Such nonlinear derivative interactions are crucial for realizing the Vainshtein mechanism.
However, we ignore for example $(\nabla\varepsilon)^{4}$ since it has less derivatives than
$(\nabla\varepsilon)^{2}(\nabla^{2}\varepsilon)^{2}$ and hence is smaller.

By doing the expansion following the above rule, we obtain the effective action of the form
\begin{eqnarray}
S^{\rm eff}=\int\D t\D^3\mathbf{x}\,{\cal L}^{\rm eff}
=\int\D t\D^3\mathbf{x}\left({\cal L}^{\rm eff}_{\rm H}+{\cal L}^{\rm eff}_{\rm beyond}\right),
\end{eqnarray}
where ${\cal L}_{{\rm H}}^{{\rm eff}}$ is the interactions included in the Horndeski class
and ${\cal L}_{{\rm beyond}}^{{\rm eff}}$ is those beyond Horndeski.
Let us first look at the effective Lagrangian derived from the Horndeski part:
\begin{eqnarray}
{\cal L}^{\rm eff}_{\rm H}
={\cal L}_{\rm H}^{(2)}
+{\cal L}_{\rm H}^{{\rm NL}} ,\label{eq:lag_H_eff}
\end{eqnarray}
where the quadratic and nonlinear terms are given by
\begin{eqnarray}
{\cal L}_{\rm H}^{(2)}&=&-a{\cal F}\Psi\nabla^2\Psi+2a{\cal G}\Psi\nabla^2\Phi
+\frac{a\eta}{2}\pi\nabla^2\pi
-2a\xi_1\Phi\nabla^2\pi+4a\xi_2\Psi\nabla^2\pi
-\frac{a^3\rho}{\mpl}\Phi\delta,
\label{H-2}
\\
{\cal L}_{\rm H}^{{\rm NL}}&=&
\frac{\mu}{a\Lambda^3}{\cal L}_3^{{\rm Gal}}+\frac{\nu}{a^3\Lambda^6}{\cal L}_4^{{\rm Gal}}
-\frac{\alpha_1}{a\Lambda^3}\Phi{\cal E}_3^{{\rm Gal}}
+\frac{\alpha_2}{a\Lambda^3}\Psi{\cal E}_3^{{\rm Gal}}
-\frac{\beta}{a^3\Lambda^6}\Phi{\cal E}_4^{{\rm Gal}}
+\frac{\gamma}{a\Lambda^3}\nabla^i\Phi\nabla^j\Psi\left(
\delta_{ij}[\Pi]-\Pi_{ij}
\right),\label{H-NL}
\end{eqnarray}
respectively.
Here, ${\cal L}^{\rm Gal}_a$ is the (spatial) Galileon Lagrangians
and ${\cal E}^{\rm Gal}_a$ are the corresponding
equations of motion derived from ${\cal L}^{\rm Gal}_{a}$:
\begin{eqnarray}
{\cal L}_3^{{\rm Gal}}&=&-\frac{1}{2}(\nabla\pi)^2[\Pi],
\\
{\cal L}_4^{{\rm Gal}}&=&-\frac{1}{2}(\nabla\pi)^2{\cal E}^{{\rm Gal}}_3,
\\
{\cal E}_3^{{\rm Gal}}&=&[\Pi]^2-[\Pi^2],
\\
{\cal E}_4^{{\rm Gal}}&=&[\Pi]^3-3[\Pi][\Pi^2]+2[\Pi^3],
\end{eqnarray}
where we have introduced the convenient notation
\begin{eqnarray}
\Pi_{ij}=\nabla_i\nabla_j\pi,\quad
\Pi^n_{ij}=\nabla_i\nabla^{k_{1}}\pi\nabla_{k_{1}}\nabla^{k_{2}}\pi
\cdots\nabla^{k_{n-1}}\nabla_j\pi,
\quad
[\Pi^n]=\delta^{ij}\Pi^n_{ij}.
\end{eqnarray}
The quintic Galileon Lagrangian,
${\cal L}_5^{{\rm Gal}}=-(\nabla\pi)^2{\cal E}_4^{{\rm Gal}}/2$,
does not appear in the effective action because it is a total divergence in three dimensions.
In Eq.~(\ref{H-2}) and~(\ref{H-NL})
we have defined the time-dependent dimensionless coefficients,
${\cal F}, {\cal G}, \eta, \xi_1, \xi_2,\mu,\nu, \alpha_1, \alpha_2, \beta, \gamma$,
which can be written in terms of $H(t)$, $\phi(t)$, and $G_a(\phi(t), X(t))$.
Their concrete form can be deduced from Ref.~\cite{Kimura:2011dc}.
For the purpose of the present paper it is sufficient to
write only some of the coefficients explicitly in terms of $G_{a}$
as summarized in Appendix A.

We assume that these coefficients are ${\cal O}(1)$ unless they vanish,
based on the underlying assumption that $G_{a}$ scale as
\begin{eqnarray}
G_{2X}\sim 1,\quad
G_{3X}\sim \Lambda^{-3},\quad
G_{4X}\sim \mpl \Lambda^{-3},\quad
G_{5X}\sim\mpl \Lambda^{-6},\quad
X\sim\mpl \Lambda^3,
\end{eqnarray}
where $\Lambda$ is another mass scale.
To explain the cosmic acceleration due to the interaction of the scalar field,
it is reasonable to assume that $\Lambda$ is
related to the present value of the Hubble rate, $H_0$, as $\Lambda^3\sim \mpl H_0^2$.
Then it is easy to see that the scaling is consistent with
the cosmological background equations.
A concrete realization of the above scaling and self-accelerating cosmology with $H\sim(\Lambda^3/\mpl)^{1/2}$
can be found for example in massive gravity~\cite{deRham:2010kj}.
The effective Lagrangian~(\ref{eq:lag_H_eff}) is the cosmological generalization
of Ref.~\cite{KNT} in the Newtonian gauge,
and the corresponding equations of motion have
already been derived in Ref.~\cite{Kimura:2011dc}.

The two Lagrangians beyond Horndeski,
$\widetilde{L}_4$ and $\widetilde{L}_5$,
generate the following new terms,
\begin{eqnarray}
{\cal L}_{\rm beyond}^{(2)}&=&\frac{4a\xi_t}{{\cal M}}\dot\Psi\nabla^2\pi,\label{beyond-2}
\\
{\cal L}_{{\rm beyond}}^{{\rm NL}}&=&\frac{2\alpha_t}{a\Lambda^3{\cal M}}\dot{\Psi}{\cal E}_3^{{\rm Gal}}
-\frac{4\alpha_\ast}{a\Lambda^3}
\nabla_i\Psi\nabla_j\pi\Pi_{ij}
-\frac{4\beta_\ast }{a^3\Lambda^6}
\nabla_i\Psi\nabla_j\pi\left([\Pi]\Pi_{ij}-\Pi^2_{ij}\right),\label{beyond-NL}
\end{eqnarray}
where
we have introduced the dimensionless parameters of ${\cal O}(1)$ (unless they vanish)
as
\begin{eqnarray}
\frac{\mpl\xi_t}{{\cal M}}&:=&\dot\phi X\left(F_4 -2H\dot\phi F_5\right),
\\
\frac{\mpl\alpha_t}{\Lambda^3{\cal M}}&:=&\dot\phi XF_5,
\\
\frac{\mpl \alpha_*}{\Lambda^3}&:=&X\left( F_4-2H\dot\phi F_5\right),
\\
\frac{\mpl \beta_*}{\Lambda^6}&:=&XF_5,
\end{eqnarray}
with ${\cal M}$ being a mass scale related to $\Lambda$ through ${\cal M}^2 =\Lambda^3/\mpl$.
Here we have assumed the scaling
\begin{eqnarray}
F_4\sim\Lambda^{-6},\quad F_5\sim\mpl \Lambda^{-6}.
\end{eqnarray}
This scaling is consistent with the cosmological background equations.
Note that the two Lagrangians also generate
the same Horndeski terms as appearing in Eqs.~(\ref{H-2}) and~(\ref{H-NL}),
which can be absorbed in the redefinition of
the corresponding coefficients.
We thus move all the Horndeski-type interactions
arising from $\widetilde{L}_4$ and $\widetilde{L}_5$ into ${\cal L}_{\rm H}^{\rm eff}$,
and define ${\cal L}_{\rm beyond}^{\rm eff}$ solely in terms of the new interactions as
${\cal L}_{\rm beyond}^{\rm eff}:={\cal L}_{\rm beyond}^{(2)}+{\cal L}_{\rm beyond}^{\rm NL}$.
From Eqs.~(\ref{beyond-2}), and~(\ref{beyond-NL})
one immediately notices that field equations beyond Horndeski involve
third derivatives of the form
$\nabla^2\partial_t\varepsilon$ and $\nabla^3\varepsilon$.
The time-derivative terms here
cannot be ignored even though one employs quasi-static approximation.
For example, $\nabla^{2}\dot\Psi$ must be retained in the equations of motion
which will be of the same order as $\dot\delta$.

One sees that only in the case where $\dot\phi\neq 0$
the new interactions~(\ref{beyond-2}) and~(\ref{beyond-NL})
participate in determining small-scale gravitational fields.
Hence, taking the limit $g_{\mu\nu}\to\eta_{\mu\nu}$, $X\to 0$,
only the Horndeski terms~(\ref{H-2}) and~(\ref{H-NL}) survive, and
one finds from the concrete expression for the coefficients that
${\cal F}={\cal G}$, $\xi_1=\xi_2$, $\alpha_1=\alpha_2$, and $\gamma=0$.
The form of the resultant effective Lagrangian exactly coincides with 
the Newtonian gauge expression of the effective theory 
of the Vainshtein mechanism from
the Horndeski theory derived in Ref.~\cite{KNT} with time-derivative terms being dropped.
On the cosmological background, however,
the above degeneracy of the coefficients
is disentangled in general.
Note that, even in the case of a Minkowski background,
one can consider a solution with $X=$ const $\neq 0$
({\em i.e.,} $\phi ={\rm const}\,\times\,t$) in a certain subclass of
the general scalar-tensor theories,
and the degeneracy of the coefficients is disentangled around such a background as well
due to the preferred time direction specified by $\nabla_\mu\phi$.
On such a background ${\cal L}_{\rm beyond}^{\rm NL}$ does not vanish.

To summarize, the total effective Lagrangian is given by
\begin{eqnarray}
{\cal L}^{\rm eff}={\cal L}_{\rm H}^{(2)}+{\cal L}_{\rm H}^{{\rm NL}}
 +{\cal L}_{\rm beyond}^{(2)}
+{\cal L}_{\rm beyond}^{{\rm NL}},
\end{eqnarray}
where the form of
each Lagrangian in the right-hand side is found in
Eqs.~(\ref{H-2}),~(\ref{H-NL}),~(\ref{beyond-2}), and~(\ref{beyond-NL}).
(Now the coefficients in
${\cal L}_{\rm H}^{(2)}$ and ${\cal L}_{\rm H}^{\rm NL}$ are composed not only of $G_{a}$
but also of $F_{a}$ due to the above-mentioned redefinition of the coefficients.)


\section{Linear density perturbations under the quasi-static approximation}

To explore the impact of the term
${\cal L}_{\rm beyond}^{(2)}$, let us focus on the
evolution of the density perturbation $\delta$
in the linear regime in this section, deferring the analysis of the nonlinear regime until Sec.~VI.
Since the matter sector is assumed to be {\em minimally coupled} to gravity,
the evolution equations for $\delta$ is
the same as the conventional one,
\begin{eqnarray}
\ddot\delta+2H\dot \delta=\frac{\nabla^2\Phi}{a^2\mpl}.\label{ev-density}
\end{eqnarray}
This equation is a consequence of
the conservation equations for the matter energy-momentum tensor,
and therefore can be derived without using the field equations
for the metric and the scalar field.
To close the evolution equation one needs to relate $\Phi$ to $\delta$,
and this is where the effect of modified gravity comes in.

In deriving the effective action on small scales in the previous section,
we made the quasi-static approximation
$(\nabla\varepsilon)^2\gg(\partial_t\varepsilon)^2$ in the action.
Working within the Horndeski theory, this approximation
results in three algebraic equations for $\nabla^2\Phi$, $\nabla^2\Psi$,
and $\nabla^2\pi$ at linear order~\cite{DeFelice:2011hq} .
We can thus write those three variables
in terms of $\delta$, and the evolution equation for $\delta$ is now closed.
Note that $\delta={\cal O}(\nabla^2\Phi)$, and hence
the time derivatives of $\delta$ are retained in Eq.~(\ref{ev-density}).
Since we have a single second-order differential equation for $\delta$,
we need two initial conditions to determine the quasi-static evolution of $\delta$
in the Horndeski theory.

Going beyond the Horndeski theory, ${\cal L}_{\rm beyond}^{(2)}$
gives rise to terms of ${\cal O}(\nabla^2\partial_t\varepsilon)$
in the field equations.
Since the interactions between the metric and the scalar field are rather complicated,
it is difficult to investigate the equations of motion in the original Jordan frame.
A simple way of dealing with this is
to define the Einstein-frame\footnote{We mean an Einstein frame
for linear perturbations as argued in Ref.~\cite{GLPV2},
where the background action may not take the Einstein-Hilbert form.} variables as
\begin{eqnarray}
\frac{\sqrt{{\cal F}}}{\cal G}\hat\Phi&:=&\Phi -2b_1\pi
-\frac{2b_2}{{\cal M}}\dot\pi,\label{def-hat-Phi}
\\
\frac{1}{\sqrt{{\cal F}}}\hat\Psi&:=&\Psi-\frac{\xi_1}{{\cal G}}\pi,
\end{eqnarray}
where
\begin{eqnarray}
b_1&:=&\frac{1}{{\cal G}}\left[
\frac{{\cal F}\xi_1}{2{\cal G}}
+\frac{(a\xi_t)\dot{}}{a{\cal M}}-\xi_2
\right],\\
b_2&:=&\frac{\xi_t}{{\cal G}}, 
\end{eqnarray}
and we have assumed that ${\cal F}>0$ and ${\cal G}\neq 0$.

In terms of $\hat\Phi$ and $\hat\Psi$, the quadratic part of the Lagrangian
in the Einstein frame is written as
\begin{eqnarray}
{\cal L}^{(2)}_{{\rm H}}+{\cal L}^{(2)}_{{\rm beyond}}
=-a\hat\Psi\nabla^2\hat\Psi
+2a\hat\Psi\nabla^2\hat\Phi
+\frac{a\hat\eta}{2}\pi\nabla^2\pi
-\frac{a^3}{\mpl}\left(\frac{\sqrt{{\cal F}}}{{\cal G}}\hat\Phi
+2b_1\pi
+\frac{2b_2}{{\cal M}}\dot\pi
\right)
\rho\delta,\label{Lag-Ein}
\end{eqnarray}
and we redefined the coefficient of the $\pi\nabla^2\pi$ term:
\begin{eqnarray}
\eta\to \hat\eta:=\eta-\frac{2{\cal F}\xi_1^2}{{\cal G}^2}+\frac{8\xi_1\xi_2}{{\cal G}}
+\frac{8\xi_t}{{\cal M}}\partial_t\left(\frac{\xi_1}{{\cal G}}\right)
-\frac{4}{a{\cal M}}\partial_t\left(
\frac{a\xi_t\xi_1}{{\cal G}}
\right).
\end{eqnarray}
We can thus remove ${\cal L}_{\rm beyond}^{(2)}$
as well as coupling between the metric perturbations and $\pi$
to go to the Einstein frame. In this frame, however,
$\pi$ couples to matter. In particular, ${\cal L}_{\rm beyond}^{(2)}$
yields the ``disformal'' piece $\dot\pi\delta$.\footnote{Links between
different scalar-tensor theories under disformal transformations are explored in
Refs.~\cite{Zumalacarregui:2012us, Bettoni:2013diz, Zumalacarregui:2013pma, GLPV2}.}
Working in the Einstein-frame Lagrangian, the kinetic term for small fluctuations
has the right sign provided that $\hat\eta >0$.
The field equations derived from the Lagrangian \eqref{Lag-Ein} are given by
\begin{eqnarray}
\hat\Psi&=&\hat\Phi,
\\
\nabla^2\hat\Psi&=&\frac{a^2\rho }{2\mpl}\frac{\sqrt{{\cal F}}}{{\cal G}}\delta,
\\
\hat\eta\nabla^2\pi&=&
\frac{2a^2\rho}{\mpl}b_1\delta
-\frac{2a^2\rho}{\mpl{\cal M}}(b_2\delta)\dot{},
\end{eqnarray}
where we have used the fact that $\partial_t\left( a^3\rho\right) =0$ for nonrelativistic matter.
Substituting the definitions \eqref{def-hat-Phi} into these equations and using the evolution equation \eqref{ev-density},
we see that $\nabla^2\Phi$ in the original Jordan frame can be written in the form
\begin{eqnarray}
\frac{\nabla^2\Phi}{a^2 \mpl}
=4\pi G_{\rm eff}\,\rho\,\delta -2\left(\frac{\dot\alpha}{\alpha}-H\right)\dot\delta,
\label{ddPhi}
\end{eqnarray}
where 
\begin{eqnarray}
&&8\pi G_{\rm eff}:=\frac{a^2}{\mpl^2\alpha^2}
\left\{
\frac{\cal F}{{\cal G}^2}+\frac{8b_1}{\hat\eta}\left( b_1-\frac{\dot b_2}{\cal M}\right)
+\frac{8ab_2}{\cal M}\partial_t\left[\frac{1}{a\hat\eta}\left( b_1-\frac{\dot b_2}{\cal M}\right)\right]
\right\}
\,,\\
&&\alpha:=\left(1+\frac{4b_2^2\rho}{\mpl^2{\cal M}^2\hat\eta}\right)^{1/2}a.\label{modd2}
\end{eqnarray}
One can easily see that the cosmological Poisson equation \eqref{ddPhi} is modified due to
the additional friction term from the scalar field, as well as the time-dependent effective gravitational coupling.
Within the Horndeski theory ($b_2=0$), it is clear that $\alpha$ reduces to $a$.
In this case, $G_{\rm eff}$ simply represents the effective gravitational coupling for
linear density perturbations,
the explicit form of which can be found in Ref.~\cite{DeFelice:2011hq}.
Combining Eqs.~(\ref{ev-density}) and~(\ref{ddPhi}),
a closed second-order differential equation for $\delta$ is obtained.
Going beyond Horndeski, it turns out from Eq.~(\ref{ddPhi}) that
the evolution equation for $\delta$ takes the form
\begin{eqnarray}
\ddot\delta+2\frac{\dot \alpha}{\alpha}
\dot\delta=4\pi G_{\rm eff}\,\rho\, \delta.\label{modd1}
\end{eqnarray}
Equations \eqref{modd2} and \eqref{modd1}
imply that, in addition to the usual change of the effective gravitational coupling,
${\cal L}_{\rm beyond}^{(2)}$ effectively modifies
the cosmic expansion rate felt by the density perturbation.
Note that as long as the stability condition for $\pi$ is imposed,
the singular point of this equation, $\alpha =0$,
is obviously avoided for $a>0$.
The evolution equation for the matter overdensity
with the same structure as Eq.~(\ref{modd1})
has also been derived in the context of a disformally coupled
scalar field~\cite{Koivisto:2012za}.

\section{Spherical overdensities}

In the previous section we have investigated the
impact of the new interaction beyond Horndeski
on the quasi-static evolution of linear density perturbations.
Now we are going to study the nonlinear terms beyond Horndeski
and their possible impact on the Vainshtein mechanism,
assuming spherical symmetry for simplicity.\footnote{The Vainshtein mechanism
away from spherical symmetry is nontrivial
and needs further investigation. See, e.g., Refs.~\cite{Hiramatsu:2012xj, Chagoya:2014fza}.}
We focus on the subclass of the theory in which $\beta=\gamma=0$.
This restriction is not only for simplicity, but also
for the following reasons:
within the Horndeski theory it is pointed out
that spherically symmetric Vainshtein solutions
are unstable against perturbations if $\beta\neq 0$~\cite{KNT},
and the metric potentials do not show the correct Newtonian behavior $\Phi, \Psi \sim r^{-1}$
on small scales on the cosmological background if $\gamma\neq 0$~\cite{Kimura:2011dc}.
In terms of the functions in the original Lagrangian of the theory,
the restriction $\beta=\gamma=0$ amounts to switching off $G_5$ and $F_5$.
It follows then that $\alpha_t=\beta_\ast=0$.
Note also that in $G_5=F_5=0$ theories the Friedmann equation
contains no cubic term in $H$,
and hence conventional cosmology is recovered
(except that the effective gravitational ``constant'' $G_{\rm cos}$
is time-dependent in general) for $H^2\gg \Lambda^3/\mpl ={\cal M}^2$.

Varying ${\cal L}^{\rm eff}$ with respect to $\Psi$, $\Phi$, and $\pi$,
one obtains the field equations valid on small scales,
which can be integrated once immediately.
The relevant equations are summarized as follows:
\begin{eqnarray}
&&
2\xi_2 x+{\cal G}y-{\cal F}z+\alpha_2 x^2+2\alpha_* x\left(rx'+x\right)
-\frac{2}{{\cal M}a^3}\partial_t\left( a^3\xi_t x\right)=0,\label{eq1}
\\
&&
{\cal G}z-\xi_1x-\alpha_1 x^2=A,\label{eq2}
\\
&&
\eta x -2\xi_1y+4\xi_2z+2\mu x^2+2\nu x^3-4\alpha_1xy+4\alpha_2xz
-4\alpha_\ast\left(rxz'+3xz\right) +
\frac{4\xi_t}{{\cal M}a^2}\partial_t\left( a^2z\right)
=0, \label{eq3}
\end{eqnarray}
where we have defined the dimensionless quantities as
\begin{eqnarray}
x(t,r):=\frac{1}{\Lambda^3}\frac{\pi'}{a^2r},
\quad
y(t,r):=\frac{1}{\Lambda^3}\frac{\Phi'}{a^2r},
\quad
z(t,r):=\frac{1}{\Lambda^3}\frac{\Psi'}{a^2r},
\quad
A(t,r):=\frac{1}{\mpl\Lambda^3}\frac{M(t,r)}{8\pi r^3},
\end{eqnarray}
with
\begin{eqnarray}
M(t,r):=\int^r_0 4\pi \bar r^2\rho (t)\,\delta (t,\bar r)\D\bar r,
\end{eqnarray}
being the enclosed mass.
Here $r$ is the comoving radial coordinate, and a prime ( ${}'$ ) indicates 
a partial derivative with respect to $r$.
Upon integration all the integration constants are set to $0$ because
$x=y=z=0=A$ at infinity.
Unlike theories within Horndeski, $F_4\neq 0$ theories generate second derivatives
in Eqs.~(\ref{eq1}) and~(\ref{eq3}), since the field equations away from the unitary gauge
are of third order in general.
Unlike the analysis of linear density perturbations,
we will not consider a disformal transformation of the metric
because the metric would no longer take the Newtonian gauge form
after such a transformation at nonlinear order.

In the region far from the source, we have $A\ll 1$ and
linearization is justified to get the solution $x\sim y\sim z ={\cal O}(A)$.
In this section, we consider the region in the vicinity of the source satisfying
$A\gg 1$, where nonlinearity, and hence possible new effects from ${\cal L}_{\rm beyond}^{\rm NL}$,
are expected to play a crucial role.
Let us now identify the appropriate solution for $A\gg 1$.
Using Eqs.~(\ref{eq1}) and~(\ref{eq2}),
one can eliminate $y$ and $z$ from Eq.~(\ref{eq3}).
Interestingly, in doing so $x'$ and $\partial_t x$ drop out from
Eq.~(\ref{eq3}), and we have the equation of the form
\begin{eqnarray}
\left[\left({\cal F}\xi_1-2{\cal G}\xi_2\right)A
-\frac{2{\cal G}^2\xi_t}{{\cal M}a^2}
\partial_t\left(\frac{a^2}{\cal G}A\right)\right]
+2\left[\kappa_1+
\left({\cal F}\alpha_1-{\cal G}\alpha_2+3{\cal G}\alpha_\ast\right)A + {\cal G}\alpha_\ast rA'
\right] x+\kappa_2 x^2- \Xi\, x^3=0,
\label{eq:x eq}
\end{eqnarray}
where
\begin{eqnarray}
\kappa_1&:=&-\frac{{\cal G}^2\eta}{4}+\frac{{\cal F}\xi_1^2}{2}-2{\cal G}\xi_1\xi_2
-\frac{{\cal G}\xi_t^2}{a{\mathcal M}} \partial_t \left(\frac{a\xi_1}{{\mathcal G}\xi_t}\right)
\\
\kappa_2&:=&3{\cal F}\alpha_1\xi_1-{\cal G}\left(
{\cal G}\mu+3\alpha_2\xi_1+6\alpha_1\xi_2-4\alpha_\ast \xi_1
\right)
-\frac{2a^4{\mathcal G}\xi_t^3}{\mathcal M} \partial_t \left(\frac{\alpha_1}{a^4{\mathcal G}\xi_t^2}\right)
\\
\Xi&:=&{\cal G}\left(4\alpha_1\alpha_2-2\alpha_1\alpha_\ast+{\cal G}\nu\right)
-2{\cal F}\alpha_1^2.
\end{eqnarray}

We look for solutions with $x\gg 1$ for $A\gg 1$, because for such solutions
$\pi'$ is suppressed due to nonlinearity so that we may expect that
GR is reproduced inside the Vainshtein radius. We find
the following approximate solution for $A\gg 1$ satisfying this requirement:
\begin{eqnarray}
x^2\simeq\frac{2\left[
\left({\cal F}\alpha_1-{\cal G}\alpha_2+3{\cal G}\alpha_\ast\right)A + {\cal G}\alpha_\ast rA'
\right]}{\Xi}+{\cal O}(A^{1/2}),\label{sol1}
\end{eqnarray}
where we have assumed that $rA'={\cal O}(A)$.
Substituting Eq.~(\ref{sol1}) into Eqs.~(\ref{eq1}) and~(\ref{eq2}), we obtain
\begin{eqnarray}
y&=&
\frac{
\left(2\alpha_2^2-2\alpha_2\alpha_\ast-12\alpha_\ast^2+{\cal F}\nu\right)A
-2\alpha_\ast^2\left(6rA'+r^2A''\right)
}{\Xi },\label{ysol1}
\\
z&=&
\frac{
\left(2\alpha_1\alpha_2+4\alpha_1\alpha_\ast
+{\cal G}\nu\right)A + 2\alpha_1\alpha_\ast rA'}
{\Xi }.\label{zsol1}
\end{eqnarray}
Using the concrete expression of the coefficients in terms of $G_4$ and $F_4$,
Eqs.~(\ref{ysol1}) and~(\ref{zsol1}) can be recast in a more suggestive form as
\begin{eqnarray}
y&=&8\pi G_N \mpl^{2} A-\frac{2\alpha_\ast^2}
{\Xi }\frac{(r^3A)''}{r},\label{ysol2}
\\
z&=&8\pi G_N\mpl^{2} A+\frac{2\alpha_1\alpha_\ast}
{\Xi }\frac{(r^3A)'}{r^2},\label{zsol2}
\end{eqnarray}
where it turns out that
\begin{eqnarray}
G_{N}=G_{\rm cos}.
\end{eqnarray}

Let us consider a density profile for which $\delta =0$ outside a certain radius $r_{\rm s}(t)$.
Following Refs.~\cite{{Narikawa:2013pjr}, KNT},
it will be appropriate to define the Vainshtein radius as
\begin{eqnarray}
r_{\rm V}(t):=\left[\frac{M(t,r_{\rm s})}{8\pi\mpl \Lambda^3}\right]^{1/3}.
\end{eqnarray}
Then, outside the overdensity ($r > r_{\rm s}$), we have
\begin{eqnarray}
A=\left(\frac{r_{\rm V}}{r}\right)^3.\label{arv}
\end{eqnarray}
Since $(r^3 A)'\propto r^2\delta =0$ outside the source,
the second terms in Eqs.~(\ref{ysol2}) and~(\ref{zsol2}) vanish for $r>r_{\rm s}$,
leading to $y=z=8\pi G_{N}\mpl^{2}A\;(=8\pi G_{\rm cos}\mpl^{2}A)$, {\em i.e.,}
\begin{eqnarray}
\frac{\Phi}{\mpl}=\frac{\Psi}{\mpl} =-\frac{G_{N}M(t,r_{\rm s})a^{2}}{r} .
\end{eqnarray}
This implies that GR is reproduced inside the Vainshtein radius and
outside the overdensity, provided that $G_{N}$
varies in time sufficiently slowly.\footnote{See Ref.~\cite{Babichev:2011iz} for the discussion on this point.}
Note in passing that $8\pi G_{N} ={\cal O}(1)\times \mpl^{-2}$
as the dimensionless coefficients have been assumed to be of ${\cal O}(1)$.

We thus have looked at the region outside the overdensity and found that
the $F_{4}$ term does not change the qualitative aspect of gravity near the source:
in the absence of $G_{5}$ (and $F_{5}$) the Vainshtein mechanism operates
to reproduce standard gravity, $\Phi=\Psi\;(\propto r^{-1})$, but the effective gravitational coupling
is time-dependent in general and given by $G_{\rm cos}$~\cite{Kimura:2011dc}.
The impact of the new terms beyond Horndeski
becomes manifest only inside the overdensity region,
where the second terms in Eqs.~(\ref{ysol2}) and~(\ref{zsol2}) no longer vanish.
When this is the case,
the gravitational attraction is
determined not only from the enclosed mass ($M \propto A$)
but also from the local energy density and its derivative ($\delta \rho \subset A'$, $\delta\rho'\subset A''$).
It is clear from Eqs.~(\ref{ysol2}) and~(\ref{zsol2}) that $y-z\neq 0$ for $(r^3A)'\neq 0$,
and hence the two metric potentials do not coincide:
\begin{eqnarray}
\Phi \neq \Psi,\quad(r<r_{\rm s}),
\end{eqnarray}
which implies that {\em the Vainshtein screening mechanism fails to operate
inside the overdensity region in the presence of the interaction beyond Horndeski.}
To estimate the difference between
the two potentials, let us consider for simplicity the case with $\rho\, \delta=$const.
In this case it is easy to see that $y-z={\cal O}(1)\times A$ and hence $(\Phi-\Psi)/\Phi={\cal O}(1)$ in general.
Breaking of the screening mechanism inside the overdensity region
could for example affect the motion of a galaxy inside a cluster
and modify structure of stars, though
one should take into account the effect of pressure in the stellar interior.
Astrophysical implications of this novel modification of gravity
are beyond the scope of the present paper and will be discussed in more detail elsewhere.

\begin{figure}[tbp]
  \begin{center}
    \includegraphics[keepaspectratio=true,height=90mm]{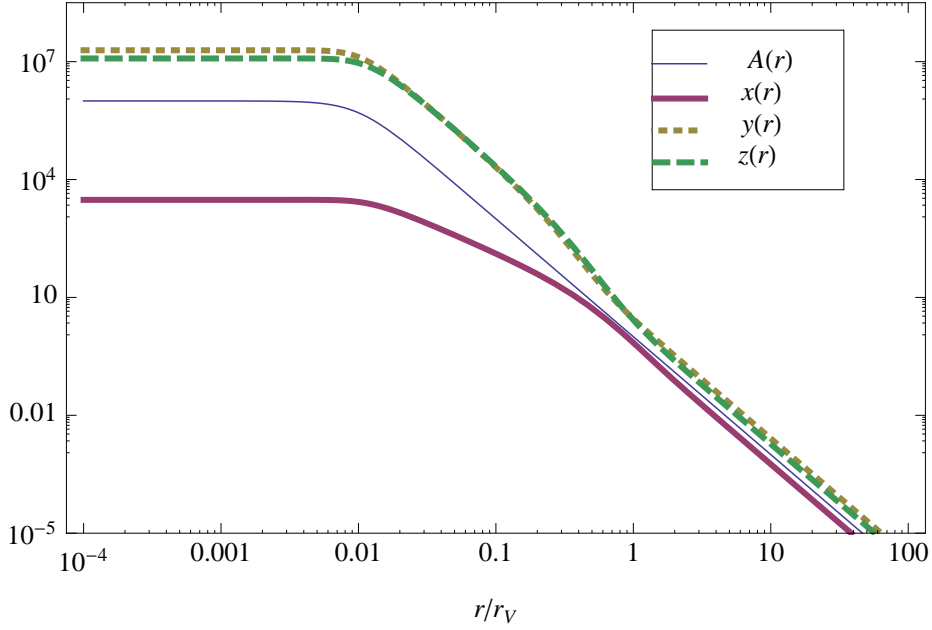}
  \end{center}
  \caption{Exact numerical solutions for $x$, $y$, $z$.
  The profile $A(r)$ is such that $A\simeq$ const for $r\lesssim 0.01\times r_{\rm V}\;(=:r_{\rm s})$
  (i.e., $r=0.01\times r_{\rm V}$ corresponds to the surface of a constant density star),
  as shown by the thin solid line. We assume that the system is static, and
  the coefficients are given by ${\cal F}=2$, ${\cal G}=1$, $\xi_1=1.5$,
  $\xi_2=1$, $\eta=0.5$, $\mu=1$,
  $\nu=0.5$, $\alpha_1=1.2$, $\alpha_2=1$, and $\alpha_\ast=-0.25$.
  The Vainshtein mechanism works well for $r_{\rm s}<r<r_{\rm V}$, giving $y/z\simeq 1.0$. However,
  for $r\gg r_{\rm V}$ we find $y/z\simeq 1.4$, while for $r\ll r_{\rm s}$ we have $y/z\simeq 1.6$,
  and hence standard gravity is not recovered there.
}%
  \label{fig:fig.eps}
\end{figure}

A numerical example is presented in Fig.~\ref{fig:fig.eps},
which agrees well with our analytic estimate.

For completeness, let us make a brief comment on another solution of Eq.~(\ref{eq:x eq}) for $A\gg 1$.
The solution is of ${\cal O}(1)$ and is given by
\begin{eqnarray}
x\simeq
\frac{-{\cal F}\xi_1/2+{\cal G}\xi_2+({\cal G}^2\xi_t/a^2{\cal M}A)\partial_t(a^2A/{\cal G})}
{{\cal F}\alpha_1-{\cal G}\alpha_2+3{\cal G}\alpha_\ast+{\cal G}\alpha_\ast rA'/A },\label{sol2}
\end{eqnarray}
where we have assumed that $rA'={\cal O}(A)$ and $\dot A/{\cal M}={\cal O}(A)$.
It can be seen immediately from Eqs.~(\ref{eq1}) and~(\ref{eq2})
that for this $x$,
\begin{eqnarray}
y=\frac{{\cal F}}{{\cal G}^2}A,\quad
z=\frac{1}{{\cal G}}A.\label{anothersol}
\end{eqnarray}
Therefore, $\Psi/\Phi = {\cal G}/{\cal F}\neq 1$ in general
even in the Vainshtein radius.\footnote{Within the Horndeski theory,
the exceptional case, ${\cal F}={\cal G}$, corresponds to $c_{t}=1$,
where $c_{t}$ is the propagation speed of gravitational waves.
In the presence of the terms beyond Horndeski, the propagation speed
is given in general by~\cite{GLPV}
\begin{eqnarray}
\frac{1}{c_t^2}=\frac{{\cal G}}{{\cal F}}+\frac{4X^2}{\mpl^2{\cal F}}
\left[
-F_4+2F_5\left(H\dot\phi+\ddot\phi\right)
\right],
\end{eqnarray}
and hence the condition ${\cal F}={\cal G}$ does not
correlate with the speed of gravitational waves.}
This property is independent of whether the region inside or outside
the overdensity is considered.

\section{Conclusions}

In this paper, we have derived 
an effective theory on small scales
from scalar-tensor theories beyond Horndeski~\cite{GLPV}.
In doing so, we have made a quasi-static approximation in the Newtonian gauge,
while keeping all the nonlinear terms which could be relevant on small scales.
We have seen that
the new interaction terms beyond Horndeski give rise to
third derivatives in the field equations.
The impact of those new terms has been investigated
in the case of $L_5=\widetilde{L}_5=0$ theories.
We have found that the linear growth of matter density perturbations
is modified not only through the time-dependent effective gravitational coupling
but also through the additional friction, which is absent in the Horndeski theory.
This is due to the disformal coupling of the scalar field to matter in the Einstein frame.
We have then investigated the nonlinear effect of the scalar-field fluctuations
that can screen the fifth force inside the Vainshtein radius.
One of the solutions outside and near the source has been shown to
reproduce the standard behavior, $\Phi=\Psi\;(\sim r^{-1})$, though the effective gravitational coupling, $G_{N}$,
is time-dependent in general on a cosmological background, as in the Horndeski theory.
In particular, $G_{N}$ coincides with the gravitational coupling in the Friedmann equation
even in the presence of the $F_{4}$ term.
However, the new interactions beyond Horndeski crucially change
the behavior of the gravitational potentials
inside the matter overdensity. The gravitational attraction
depends not only on the enclosed mass but also on the local matter energy density,
and $\Phi$ and $\Psi$ no longer coincide, implying that GR is not recovered inside the source.

We have made the quasi-static approximation for scalar perturbations
throughout the paper. This hindered us from investigating
the stability of the spherical solutions in Sec.~VI.
It would be interesting to study how the new terms beyond Horndeski
change the stability {\em e.g.} of stars.

\acknowledgments
This work was supported in part by JSPS Grant-in-Aid for Young
Scientists (B) No.~24740161 (T.K.), 
the JSPS Research Fellowship for Young Scientists Nos.~269337 (Y.W.)
and 259800 (D.Y.),
and the Munich Institute for Astro- and Particle Physics (MIAPP)
of the DFG cluster of excellence ``Origin and Structure of the Universe'' (Y.W.).

\appendix

\section{Explicit expressions for the background quantities and the coefficients in the effective action}

\subsection{Background quantities in the Horndeski theory}

The cosmological background evolution in the Horndeski theory 
is given in terms of
the gravitational analogue of the energy density ${\cal E}_{\rm H}$, the isotropic pressure ${\cal P}_{\rm H}$,
and the
current $J_{\rm H}$ of the scalar field $\phi$~\cite{GG}:
\begin{eqnarray}
{\cal E}_{\rm H} := \sum^5_{a=2}{\cal E}_a,
\quad
{\cal P}_{\rm H} := \sum^5_{a=2}{\cal P}_a,
\end{eqnarray}
and
\begin{eqnarray}
J_{\rm H} &:=& \dot\phi G_{2X} +6HXG_{3X} -2\dot\phi G_{3\phi} +6H^2\dot\phi (G_{4X}+2XG_{4XX}) -12HXG_{4X\phi} \nonumber\\
&&+2H^3X(3G_{5X}+2XG_{5XX}) -6H^2\dot\phi (G_{5\phi}+XG_{5X\phi}),
\end{eqnarray}
where
\begin{eqnarray}
{\cal E}_2 &=& 2XG_{2X}-G_2,\\
{\cal E}_3 &=& 6HX\dot{\phi}G_{3X}-2XG_{3\phi},\\
{\cal E}_4 &=& -6H^2G_4+24H^2X(G_{4X}+XG_{4XX})-12HX\dot\phi G_{4X\phi}-6H\dot{\phi} G_{4\phi},\\
{\cal E}_5 &=& 2H^3X\dot{\phi} (5G_{5X}+2XG_{5XX})-6H^2X(3G_{5\phi}+2XG_{5X\phi}),\\
{\cal P}_2 &=& G_{2},\\
{\cal P}_3 &=& -2X(G_{3\phi}+\ddot\phi G_{3X}),\\
{\cal P}_4 &=& 2(3H^2+2\dot{H})G_4 -4(3H^2X+H\dot{X}+2\dot{H}X)G_{4X} -8HX\dot{X}G_{4XX} \nonumber\\
&&-4X(\ddot\phi -2H\dot\phi) G_{4X\phi} +2H(\ddot{\phi}+2H\dot\phi) G_{4\phi} +4XG_{4\phi\phi},\\
{\cal P}_5 &=& -2X(2H^3\dot{\phi}+2H\dot{H}\dot\phi+3H^2\ddot\phi)G_{5X} -4H^2X^2\ddot\phi G_{5XX} \nonumber\\
&&+4HX(\dot{X}-HX)G_{5X\phi} +2[2(HX){\bf\dot{}} +3H^2X]G_{5\phi} +4HX\dot\phi G_{5\phi\phi}.
\end{eqnarray}
The equation of motion for the scalar field is given by $(a^3 J_{\rm H}){\bf\dot{}}=a^3P_{{\rm H}\phi}$
where
\begin{eqnarray}
P_{{\rm H}\phi} &:=&
 G_{2\phi} -2X(G_{3\phi\phi}+\ddot\phi G_{3X\phi}) +6(2H^2+\dot{H})G_{4\phi} +6H(\dot{X} +2HX)G_{4X\phi} \nonumber\\
&&-6H^2XG_{5\phi\phi} +2H^3X\dot\phi G_{5X\phi}.
\end{eqnarray}

\subsection{Coefficients in the effective action}

We basically follow the notations in Refs.~\cite{Narikawa:2013pjr, KNT}
rather than those in Ref.~\cite{Kimura:2011dc}. Here we summarize the
relations between the coefficients in the effective action and
those of the equations of motion derived in Ref.~\cite{Kimura:2011dc}.
In the absence of the terms beyond Horndeski, the coefficients in Eqs.~(\ref{H-2}) and~(\ref{H-NL})
are given by
\begin{eqnarray}
&&
\mpl^2{\cal F}={\cal F}_T,\quad
\mpl^2{\cal G}={\cal G}_T,\quad
\mpl\eta = -\frac{H^2A_0}{X},\quad
\mpl\xi_1 = -\frac{H}{\dot\phi}A_2,\quad
\mpl\xi_2 = \frac{H}{2\dot\phi} A_1,
\nonumber\\ &&
\frac{\mpl \alpha_1}{\Lambda^3}=-\frac{B_2}{2X},
\quad
\frac{\mpl \alpha_2}{\Lambda^3} =\frac{B_1}{2X},
\quad
\frac{\mpl\beta}{\Lambda^6}= -\frac{C_1}{3\dot\phi XH},
\quad
\frac{\mu}{\Lambda^3}=-\frac{HB_0}{\dot\phi X},
\quad
\frac{\nu}{\Lambda^6}=\frac{C_0}{2X^2},
\quad
\frac{\mpl^2\gamma}{\Lambda^3}=-\frac{2B_3}{\dot\phi H},\qquad
\end{eqnarray}
where $A_0$, $A_1$, ... in the right-hand sides are found in Appendix A of Ref.~\cite{Kimura:2011dc}.
The Lagrangians $\widetilde{L}_4$ and $\widetilde{L}_5$ generate the Horndeski terms
as well as the new terms~(\ref{beyond-2}) and~(\ref{beyond-NL}).
Due to this, the above coefficients are redefined as
\begin{eqnarray}
\mpl^2{\cal F}&\to &\mpl^2{\cal F},
\\
\mpl^2{\cal G}&\to &\mpl^2{\cal G},
\\
\mpl\eta &\to&\mpl\eta + 4\left(10H^2X+5\dot HX+6H\dot X\right)F_4
+4X\left(5H^2X+2\dot H X+9H\dot X\right)F_{4X}+8HX^2\dot XF_{4XX}
\nonumber\\&&
+12HX\dot\phi F_{4\phi}+8HX^2\dot\phi F_{4\phi X}
-8HX\left(5H^2\dot\phi+6\dot H \dot\phi+10H \ddot\phi\right)F_5
-8HX^2\left(2H^2\dot\phi+2\dot H\dot\phi + 11 H\ddot\phi\right)F_{5X}
\nonumber\\&&
-16H^2X^3\ddot\phi F_{5XX}-32H^2X^2F_{5\phi}-16H^2X^3F_{5\phi X},
\\
\mpl\xi_1 &\to& \mpl\xi_1-2HX\dot\phi\left( 5F_4+2XF_{4X}\right)+4H^2X^2\left(7F_5+2XF_{5X}\right),
\\
\mpl \xi_2&\to& \mpl \xi_2,
\\
\frac{\mpl \alpha_1}{\Lambda^3} &\to&\frac{\mpl \alpha_1}{\Lambda^3}
+X\left( 5F_4+2XF_{4X}\right)-2HX\dot\phi\left(7F_5+2X F_{5X}\right),
\\
\frac{\mpl \alpha_2}{\Lambda^3} &\to&\frac{\mpl \alpha_2}{\Lambda^3}
+XF_4-2HX\dot\phi F_5,
\\
\frac{\mpl \beta}{\Lambda^6}&\to& \frac{\mpl \beta}{\Lambda^3}+\frac{2X}{3}\left(7F_5+2XF_{5X}\right),
\\
\frac{\mu}{\Lambda^3}&\to&\frac{\mu}{\Lambda^3}-2\left(\ddot\phi+5H\dot\phi\right)
F_{4}-5X\left(\ddot\phi+H\dot\phi\right) F_{4X}-2X^2\ddot\phi F_{4XX}+XF_{4\phi}-2X^2F_{4\phi X}
\nonumber\\&&+2\left(5H^2X+2H\dot X +14\dot H X\right)F_5
+2\left(2H^2X+2\dot H X+11H\dot X\right)XF_{5X}
\nonumber\\&&
+4HX^2\dot XF_{5XX}+8HX\dot \phi F_{5\phi}+4HX^2F_{5\phi X},
\\
\frac{\nu}{\Lambda^3}&\to&\frac{\nu}{\Lambda^3}+2F_4+XF_{4X}
-\frac{20}{3}\ddot\phi F_5-\frac{22}{3}X\ddot\phi F_{5X}-\frac{4}{3}X^2\ddot\phi F_{5XX}
-\frac{8}{3}XF_{5\phi}-\frac{4}{3}X^2F_{5\phi X},
\\
\frac{\mpl^2 \gamma}{\Lambda^3}&\to& \frac{\mpl^2 \gamma}{\Lambda^3}.
\end{eqnarray}
We investigate the theories with $G_5=F_5=0$ in the main text.
In this case it can be seen that $\beta = \gamma =0$. One may also notice that
the following relation holds:
\begin{eqnarray}
\nu\left({\cal F}-{\cal G}\right)+2(\alpha_2-\alpha_1)(\alpha_2-\alpha_\ast)=0.\label{relation4}
\end{eqnarray}



\end{document}